# ChMusic: A Traditional Chinese Music Dataset for Evaluation of Instrument Recognition

Xia Gong

School of Music, Shandong University of Technology, Zibo, China, sdlggx@sdut.edu.cn

Yuxiang Zhu

No.2 High School (Baoshan) of East China Normal University, Shanghai, China, 3447535189@qq.com

Haidi Zhu

Shanghai Institute of Microsystem and Information Technology, Chinese Academy of Sciences, Shanghai, China, hdzhu@mail.sim.ac.cn

Haoran Wei

Department of Electrical and Computer Engineering, University of Texas at Dallas Richardson, USA, haoran.wei@utdallas.edu

**Musical instruments recognition is a widely used application for music information retrieval. As most of previous musical instruments recognition dataset focus on western musical instruments, it is difficult for researcher to study and evaluate the area of traditional Chinese musical instrument recognition. This paper collected a traditional Chinese music dataset for training model and performance evaluation, named ChMusic. This dataset is free and publicly available, 11 traditional Chinese musical instruments and 55 traditional Chinese music excerpts are recorded in this dataset. Then an evaluation standard is proposed based on ChMusic dataset. With this standard, researchers can compare their results following the same rule, and results from different researchers will become comparable.**

## 1 INTRODUCTION

Musical instruments recognition is an important and fascinating application for music information retrieval(MIR). Deep learning is great tool for image processing [1,2,3], video processing [4,5,6], natural language processing [7,8,9], speech processing [10,11,12] and music processing [13,12,15] applications and has demonstrated better performance compared with previous approaches. While deep learning based approaches rely on proper datasets to train model and evaluate performance.

Though there are already several musical instruments recognition dataset, most of these datasets only covers western musical instruments. For examples, Good-sounds dataset [16] has 12 musical instruments consisting of flute, cello, clarinet, trumpet, violin, sax alto, sax tenor, sax baritone, sax soprano, oboe, piccolo and bass. Another musical instruments recognition dataset called [17] has 20 musical instruments consisting of accordion, banjo, bass, cello, clarinet, cymbals, drums, flute, guitar, mallet percussion, mandolin, organ, piano, saxophone, synthesizer, trombone, trumpet, ukulele, violin, and voice. Very few attention has been made to other culture's musical instruments recognition. Even for the very limited researches base on Chinese musical instruments recognition [18,19], dataset used for these research are not publicly available. Without an open access Chinese musical instruments dataset, Another critical problem is that researchers cannot evaluate their model performance by a same standard, so results reported from their papers are not comparable.

To deal with the problems mentioned above, a traditional Chinese music dataset, named ChMusic, is proposed to help training Chinese musical instruments recognition models and then conducting performance evaluation. So the major contributions of this paper can be concluded as:

1) Propose a traditional Chinese music dataset for Chinese traditional instrument recognition, named ChMusic.

2) Come up with an evaluation standard to conduct Chinese traditional instrument recognition on ChMusic dataset.

3) Propose baseline approach for Chinese traditional instrument recognition on ChMusic dataset, code for this baseline is https://github.com/HaoranWeiUTD/ChMusic

The rest of the paper is organized as follows: Section 2 covers details of ChMusic dataset. Then, the corresponding evaluation standard to conduct Chinese traditional instrument recognition on ChMusic dataset are described in Section 3. Baseline for Musical Instrument Recognition on ChMusic Dataset is proposed in Section 4. Finally, the paper is concluded in Section 5.

## 2 CHMUSIC DATASET

ChMusic is a traditional Chinese music dataset for training model and performance evaluation of musical instrument recognition. This dataset cover 11 musical instruments, consisting of Erhu, Pipa, Sanxian, Dizi, Suona, Zhuiqin, Zhongruan, Liuqin, Guzheng, Yangqin and Sheng, the indicating numbers and images are shown on Figure 1 respectively.

Each musical instrument has 5 traditional Chinese music excerpts, so there are 55 traditional Chinese music excerpts in this dataset. The name of each music excerpt and the corresponding musical instrument is shown on Table 1 Each music excerpt only played by one musical instrument in this dataset.

| Number | 1 | 2 | 3 | 4 |
|---|---|---|---|---|
| Name | Erhu | Pipa | Sanxian | Dizi |
| Image | 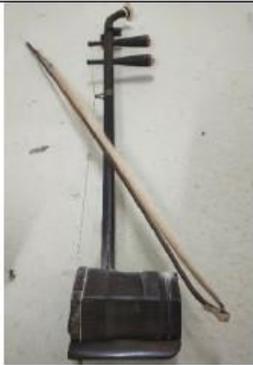 | 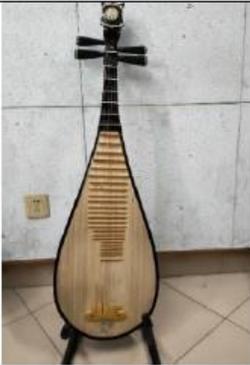 | 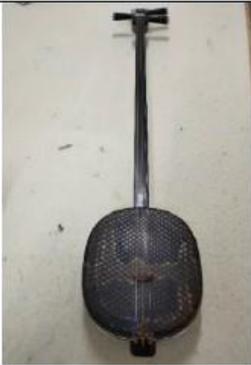 | 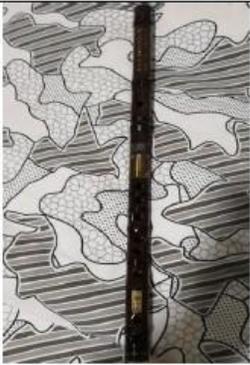 |
| Number | 5 | 6 | 7 | 8 |
| Name | Suona | Zhuiqin | Zhongruan | Liuqin |
| Image | 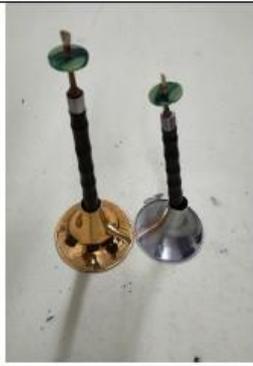 | 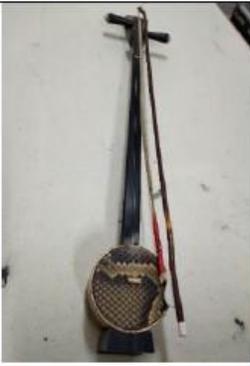 | 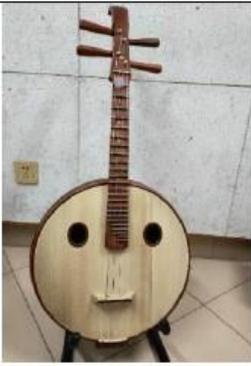 | 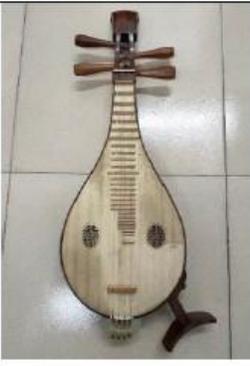 |
| Number | 9 | 10 | 11 | |
| Name | Guzheng | Yangqin | Sheng | |
| Image | 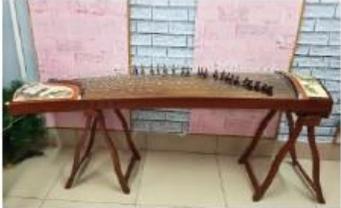 | 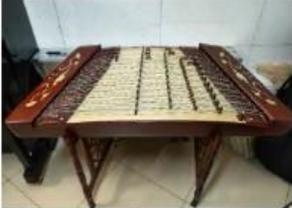 | 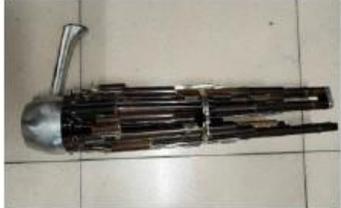 | |

Figure 1. Example of a traditional Chinese instrument.

Each music excerpt is saved as a .wav file. The name of these files follows the format of "x.y.wav", "x" indicates the instrument number, ranging from 1 to 11, and "y" indicates the music number for each instrument, ranging from 1 to 5. These files are recorded by dual channel, and has sampling rate of 44100 Hz. The duration of these music excerpts are between 25 to 280 seconds.

This ChMusic dataset can be download from Baidu Wangpan by link: pan.baidu.com/s/13e-6GnVJmC3tcwJtxed3-g with passwords: xk23, or from google drive: drive.google.com/file/d/1rfbXpkYEUGw5h_CZJtC7eayYemeFMzij/view?usp=sharing

## 3  EVALUATION OF MUSICAL INSTRUMENT RECOGNITION ON CHMUSIC DATASET

The diagram of musical instrument recognition can be simplified as Figure 2. Figure 2 consisting of two stages, corresponding to training stage and testing stage respectively. Training data is used for training stage to train a musical instrument recognition model. Then this model is tested by testing data during testing stage.

Table 1: Musics Played By Traditional Chinese Instruments

| Instrument Name | Music Number | Music Name |
|---|---|---|
| Erhu | 1 | "Ao Bao Xiang Hui" |
|  | 2 | "Mu Yang Gu Niang" |
|  | 3 | "Xiao He Tang Shui" |
|  | 4 | "Er Xing Qian Li" |
|  | 5 | "Yu Jia Gu Niang" |
| Pipa | 1 | "Shi Mian Mai Fu" excerpts |
|  | 2 | "Su" excerpts |
|  | 3 | "Zhu Fu" excerpts |
|  | 4 | "Xian Zi Yun" excerpts |
|  | 5 | "Ying Hua" |
| Sanxian | 1 | "Kang Ding Qing Ge" |
|  | 2 | "Ao Bao Xiang Hui" |
|  | 3 | "Gan Niu Shan" |
|  | 4 | "Mao Zhu Xi De Hua Er Ji Xin Shang" |
|  | 5 | "Nan Ni Wan" |
| Dizi | 1 | "Hong Mei Zan" |
|  | 2 | "Shan Hu Song" |
|  | 3 | "Wei Shan Hu" |
|  | 4 | "Xiu Hong Qi" |
|  | 5 | "Xiao Bai Yang" |
| Suona | 1 | "Hao Han Ge" |
|  | 2 | "She Yuan Dou Shi Xiang Yang Hua" |
|  | 3 | "Yi Zhi Hua" |
|  | 4 | "Huang Tu Qing" excerpts |
|  | 5 | "Liang Zhu" excerpts |
| Zhuiqin | 1 | "Jie Nian" excerpt 1 |
|  | 2 | "Jie Nian" excerpt 2 |
|  | 3 | "Hong Sao" excerpt |
|  | 4 | "Jie Nian" excerpt 3 |
|  | 5 | "Zi Mei Yi Jia" excerpt |
| Zhongruan | 1 | "Yun Nan Hui Yi" excerpt 1 |
|  | 2 | "Yun Nan Hui Yi" excerpt 2 |
|  | 3 | "Yun Nan Hui Yi" excerpt 3 |
|  | 4 | "Yun Nan Hui Yi" excerpt 4 |
|  | 5 | "Yun Nan Hui Yi" excerpt 5 |
| Liuqin | 1 | "Chun Dao Yi He" excerpt 1 |
|  | 2 | "Yu Ge" excerpt |
|  | 3 | "Chun Dao Yi He" excerpt 2 |
|  | 4 | "Yu Hou Ting Yuan" excerpt |
|  | 5 | "Jian Qi" excerpt |
| Guzheng | 1 | "Lin Chong Ye Ben" excerpt 1 |
|  | 2 | "Zhan Tai Feng" excerpt |
|  | 3 | "Yu Zhou Chang Wan" |
|  | 4 | "Lin Chong Ye Ben" excerpt 2 |
|  | 5 | "Han Jiang Yun" |
| Yangqin | 1 | "Zuo Zhu Fa Lian Xi Qu" |
|  | 2 | "Fen Jie He Xian Lian Xi Qu" |
|  | 3 | "Dan Yin Yu He Yin Jiao Ti Lian Xi Qu" |
|  | 4 | "Lun Yin Lian Xi Qu" |
|  | 5 | "Da Qi Luo Gu Qing Feng Shou" |
| Sheng | 1 | "Gua Hong Deng" excerpt |
|  | 2 | "Qin Wang Po Zhen Yue" excerpt |
|  | 3 | "Shui Ku Yin Lai Jin Feng Huang" |
|  | 4 | "Shan Xiang Xi Kai Feng Shou Lian" |
|  | 5 | "Xi Ban Ya Dou Niu Wu Qu" |

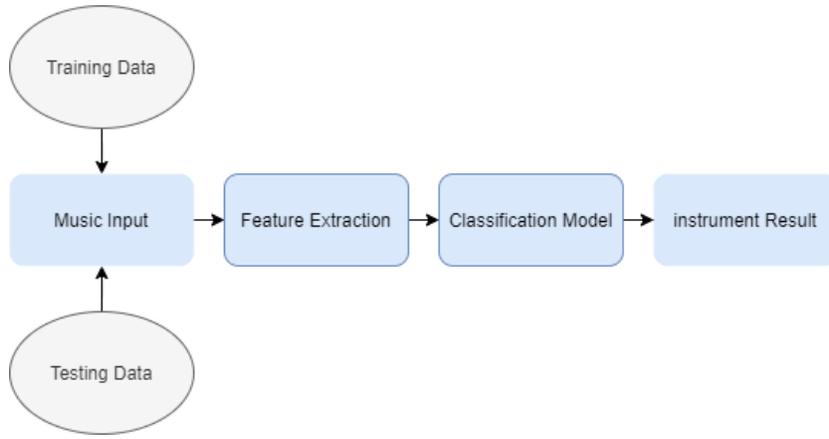

Figure 2. Diagram of musical instrument recognition.

The widely used features for musical instrument recognition include linear predictive cepstral coefficient (LPCC) [20] and mel-scale frequency cepstral coefficients (MFCC) [21] . The commonly used classification models include k-nearest neighbors (KNN) model [18] , Gaussian mixture model(GMM)[22] , support vector machine (SVM) model[23], hidden markov model (HMM) [24] and deep learning based models [25] .

In ChMusic Dataset, music with number 1 to 4 of each instrument are used as training dataset, music with number 5 of each instrument are used as testing dataset. Every music is cut into 5 seconds clips without overlap, the ending part of each music which is shorter than 5 seconds is discarded. Each 5 seconds clips need to make a prediction for musical instrument classification result.

$$Accuracy = \frac{CorrectClassifiedClips}{AllClipsFromTestingData} \qquad (1)$$

Classification accuracy will be the evaluation metrics for musical instrument recognition on ChMusic Dataset. And confusion matrix is also recommended to present the result.

## 4 BASELINE FOR MUSICAL INSTRUMENT RECOGNITION ON CHMUSIC DATASET

Baseline for musical instrument recognition on ChMusic dataset is described in this section. Machine learning [26, 27, 28, 29, 30, 31] and deep learning [32, 33, 34, 35, 36, 37, 38, 39, 40] are useful tools to deal with classification and recognition related problems [41, 42, 43, 44, 45, 46, 47, 48, 49, 50], baseline approach is conducted by machine learning approach. Twenty dimensions MFCCs are extracted from each frame, and KNN is used for frame level data training and testing. As each test clip have 5 minutes in duration, a majority voting is conducted to get final classification result from frame-level results within the 5 minutes test clip.

KNN models with various K value are compared. Accuracy for KNN models with each K value are presented in Table 2 then confusion matrix for K=3, K=5, K=9 and K=15 are presented in Figure 3, Figure 4, Figure 5 and Figure 6 respectively.

Table 2: Accuracy for Knn Models with Various K Value

| Model | Accuracy(%) |
| --- | --- |
| KNN (K=3) | 93.57 |
| KNN (K=5) | 93.57 |
| KNN (K=9) | 93.57 |
| KNN (K=15) | 94.15 |

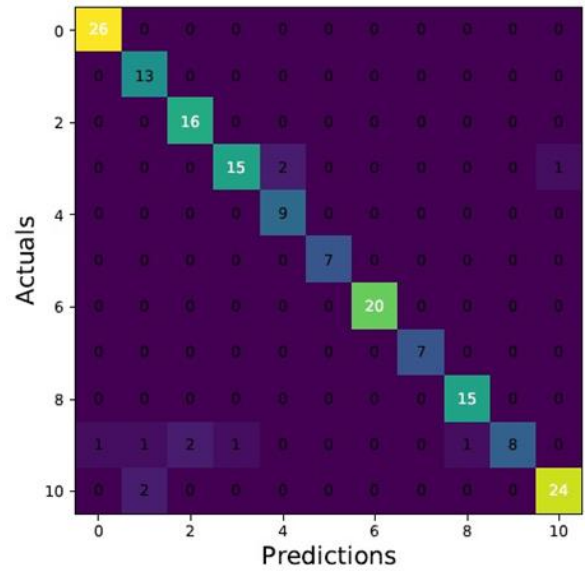

Figure 3. Confusion matrix of KNN model with K=3.

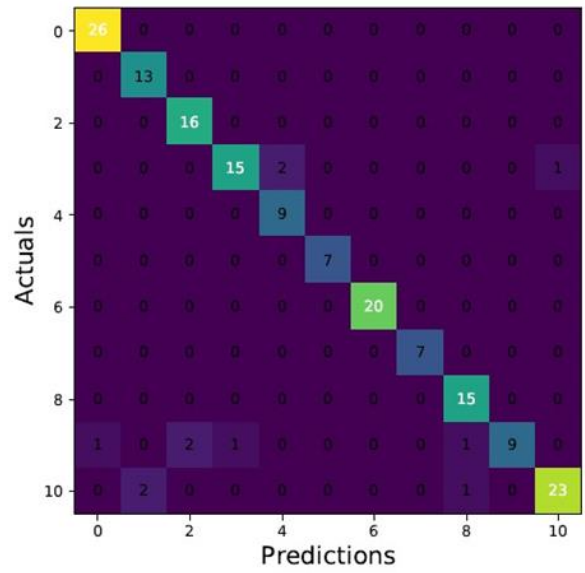

Figure 4. Confusion matrix of KNN model with K=5.

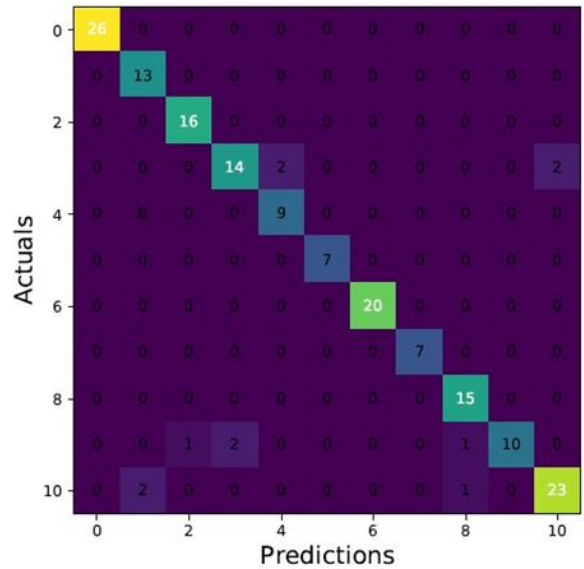

Figure 5. Confusion matrix of KNN model with K=9.

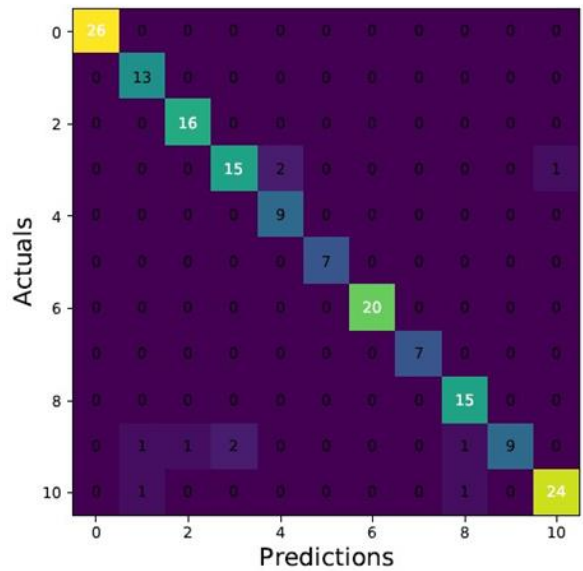

Figure 6. Confusion matrix of KNN model with K=15.

## 5 CONCLUSION

Musical instruments recognition is an interesting application for music information retrieval. While most of previous musical instruments recognition dataset focus on western musical instruments, very few works have been conducted on the area of traditional Chinese musical instrument recognition. This paper proposes a traditional Chinese music dataset for training model and performance evaluation, named ChMusic. With this dataset, researchers can download it for free. Eleven traditional Chinese musical instruments and 55 traditional Chinese music excerpts are covered in this dataset. Then an evaluation standard is proposed based on ChMusic dataset. With this standard, researchers can compare their results following the same rule, and results from different researchers will become comparable.

## 6 ACKNOWLEDGMENT

Thanks for Zhongqin Yu, Ziyi Wei, Baojin Qin, Guangyu Zhao, Qiuxiang Wang, Xiaoqian Zhang, Jiali Yao, Zheng Gao, Ke Yan, Menghao Cui and Yichuan Zhang for playing musics and helping to collect this ChMusic dataset.